\documentstyle[preprint,prc,aps]{revtex}

\begin{document}
\draft

\title{$\eta\prime$ meson production in proton-proton collisions}

\author{K. Nakayama$^{a,b}$, H. F. Arellano$^c$, J. W. Durso$^{a,d}$, and
J. Speth$^a$}

\address{$^a$Institut f\"ur Kernphysik, Forschungszentrum-J\"ulich,
D-52425, J\"ulich, Germany \\ 
$^b$Department of Physics and Astronomy, University of Georgia, Athens, 
GA 30602, USA \\
$^c$Departamento de F\'{\i}sica, Facultad de Ciencias F\'{\i}sicas y
Matem\'aticas, Universidad de Chile, Blanco Encalada 2008, Santiago, Chile \\
$^d$Physics Department, Mount Holyoke College, South Hadley, MA 01075, USA}

\maketitle

\begin{abstract}
The $pp\rightarrow pp\eta^\prime$ reaction is investigated within a relativistic 
meson-exchange model of hadronic interactions. We explore the role of nucleonic
and mesonic, as well as the $N^*$ resonance currents, in producing $\eta^\prime$ 
mesons. In order to learn more about the production mechanisms, new measurements 
in the energy region far from the threshold are required. 
\end{abstract}


\newpage

\section{ Introduction}

With the advent of particle accelerators in the few $GeV$ energy region, 
heavy meson production in hadronic collisions has attracted increasing 
attention in the past few years. In particular, heavy meson production
in nucleon-nucleon ($NN$) collisions at near-threshold energies is of 
special interest, not only because it is suited for extracting information 
on a few lowest-order multipole amplitudes, but also because it is considered to
provide important information on the short distance behavior of the $NN$
interaction. Due to the large momentum transfer between the initial 
and final nucleons, these reactions at near-threshold energies necessarily 
probe the $NN$ interaction at short distances. In the present work 
we concentrate on the $\eta^\prime$ meson production in proton-proton 
($pp$) collisions. 

Among heavy mesons, the $\eta^\prime$ meson is of particular interest for various 
reasons. The $\eta^\prime$ meson is thought to couple 
strongly to gluons via the QCD anomaly coupling $\eta^\prime \rightarrow g + g$ 
\cite{Gluons}. Also, an unexpectedly large branching ratio measured recently for the 
inclusive decay of beauty particles, $B \rightarrow \eta ^{\prime }+X$ \cite{CLEO} 
has been interpreted as possible evidence for the strong coupling of $\eta^\prime$ 
meson to the gluonic components\cite{Gluonic}. It would then be conceivable that the 
$pp \rightarrow pp\eta^\prime$ reaction might probe the gluon content of the 
$\eta^\prime$ meson via its coupling to gluons emitted from the quarks exchanged 
between two interacting nucleons. This mechanism would be complementary to the vector 
meson-exchange current mechanism for producing $\eta^\prime$ mesons.

One of the properties of the $\eta ^{\prime }$ meson of 
extreme importance is its yet-poorly-known coupling strength to the nucleon. 
This has attracted much attention in connection with the so-called 
''nucleon-spin crisis'' in polarized deep inelastic lepton scattering 
\cite{EMC88}. The $NN\eta ^{\prime }$ coupling constant, $g_{NN\eta ^{\prime }}$, 
may be related (through the axial vector coupling using the Goldberger-Treiman 
relation) to the quark helicity contribution to the spin of the proton 
\cite{Efremov}. Therefore one can argue that $g_{NN\eta ^{\prime }}$ tells us 
about the total spin of the nucleon carried by its constituents; conversely, the 
quark contribution to the spin of the nucleon would tell us about 
$g_{NN\eta ^{\prime }}$. So far there is no direct experimental measurement of 
$g_{NN\eta ^{\prime }}$. Recently, cross sections for the $pp \rightarrow 
pp\eta ^{\prime }$ reaction near threshold have been measured by the COSY-11 
\cite{Moskal} and SPESIII \cite{Hibou} collaborations. This reaction may offer 
an opportunity to determine this coupling from a direct emission process of 
$\eta^\prime$ by a proton. Of course, other production mechanisms, such as 
meson exchange and nucleon resonance currents, must be taken into account 
before a quantitative determination of $g_{NN\eta^\prime}$ is possible. The major
aim of the present work is to explore the roles of various production mechanisms.

The theory of $\eta ^{\prime }$ production in $pp$ collisions is still in its 
early stage of development \cite{Sibirtsev,Bernard,Gedalin,Baru}. In this 
work we investigate this reaction using a relativistic meson-exchange model of 
hadronic interactions. In section 2 we outline the formalism for calculating the 
production amplitude. The final state interaction is known to play a crucial role
in the production of particles near threshold energies in $NN$ collisions
\cite{Watson}; therefore our formalism includes the $pp$ final state interaction 
explicitly. The Coulomb correction in the $pp$ final state interaction is also known 
to be important when these protons have small energies \cite{Miller,Herrmann1}. In the
present work it is treated exactly and reduces the calculated cross section by as much 
as a factor of two for energies close to threshold. The initial state interaction is 
taken into account via a reduction factor \cite{Hanhart} determined from the available 
phase shifts and inelasticities in the threshold incident energy region. In section 3 our 
$\eta ^{\prime }$ meson production currents are constructed. We consider the nucleonic, 
mesonic and resonance currents. The roles of these currents are explored in section 4. 
Our results are summarized in Section 5.

\section{Formalism}

The formalism used in the present work is essentially the same as that
employed in Refs.\cite{Nak1,Nak2} for studying the production of vector mesons.
It is based on a relativistic meson-exchange model of hadronic interactions
in which the transition amplitude is calculated in the Distorted Wave Born 
Approximation, taking explicitly into account effects of the final state 
interaction. We write the transition amplitude describing the 
$p+p\rightarrow p+p+\eta^{\prime }$ process as
\begin{equation}
M = < \phi_f | (T^{(-)\dagger}_f iG_f + 1) J | \phi_i > \ ,
\label{amplitude}
\end{equation}
where $\phi _{i,f}$ denotes the four-component unperturbed $pp$
wave function in the initial ($i$) and final ($f$) state. $T_{f}^{(-)}$ is
the final state $pp$ T-matrix. $G_{f}$ stands for the two-nucleon propagator
and $J$ is the $\eta ^{\prime }$-emission current defined in the next
section.

The T-matrix used in our calculation is generated by solving a
three-dimensional reduced Bethe-Salpeter equation (the Blankenbecler-Sugar
equation) for a relativistic one-boson-exchange $NN$ potential $V$, i.e.,
\begin{equation}
T = V + ViG_{BBS}T \ ,
\label{scateqn}
\end{equation}
where $G_{BBS}$ denotes the Blankenbecler-Sugar (BBS) two-nucleon propagator. 
In this work we employ a slightly modified version \cite{Haidenbauer} of the 
Bonn B $NN$ model as defined in Table A.1 of Ref.\cite{MHE87} for constructing 
the potential $V$. This modification has been made in order to reproduce the $pp$ 
low-energy parameters rather than the $pn$ low-energy parameters while preserving 
other features of the Bonn B interaction model for describing $pp$ scattering 
\cite{Haidenbauer}. It should also be mentioned that each nucleon-nucleon-meson 
($NNM$) vertex in the $NN$ potential is modified by a form factor of either 
monopole or dipole form. We refer to \cite{MHE87} for further details. 
Furthermore we note that the two-nucleon propagator $G_{f}$ appearing in 
Eq.(\ref{amplitude}) is, for consistency, also chosen to be the BBS propagator, 
$G_{f}=G_{f;BBS}$.

In Eq.(\ref{amplitude}) the initial state interaction is neglected. For heavy 
meson production such as $\eta ^{\prime }$, its effect on the production cross
sections near threshold may be taken into account by multiplying the calculated
cross sections using Eq.(\ref{amplitude}) by a factor \cite{Hanhart}
\begin{equation}
\lambda_\alpha = \eta_\alpha \cos ^2(\delta_\alpha ) + 
              \frac{1}{4}[1-\eta_\alpha]^2
\label{isifactor}
\end{equation}
where $\alpha$ stands for the quantum numbers specifying the corresponding initial 
$NN$ state and $\delta _{\alpha }$ and $\eta _{\alpha }$ denote the corresponding
$NN$ phase-shift and inelasticity, respectively, at the nucleon incident energy. 
Although this reduction factor is only meant to account for the gross effect of the 
initial state interaction, it works well when confronted with a calculation where 
the initial state interaction has been treated explicitly \cite{lee}. In the near 
threshold energy region the relevant initial $pp$ state is $\alpha=\ ^3P_0$. For 
$\eta^\prime$ production the threshold incident energy is about $T_{lab}=2.404~GeV$. 
Using the values $\delta _{(^3P_0)}=-50{{}^\circ}$ and $\eta _{(^3P_0)}=0.75$ from 
a partial wave analysis \cite{said}, we obtain the reduction factor of 
$\lambda_\alpha=0.33$ for the cross section. An alternative way to account 
for effects of the initial state interaction is to absorb these effects in the 
form factors at the $\eta ^{\prime }$ meson production vertices as has been done 
in Refs.\cite{Nak1,Nak2}. We choose the former method in the present work.

\section{Production Currents}

Within our hadronic model of strong interactions, the $\eta ^{\prime }$-emission 
current $J$ in Eq.(\ref{amplitude}) consists of a sum of the baryonic and mesonic
currents. The baryonic current is further divided into the nucleonic and nucleon 
resonance ($N^*$) currents, so that the total current is given by
\begin{equation}
J = J_{nuc} + J_{res} + J_{mec} \ .
\label{curr}
\end{equation}
The individual currents are illustrated diagrammatically in Fig.~~\ref{fig1}. In the 
following subsections we construct each of these currents.

\subsection{The Nucleonic Current}

The nucleonic current is defined as
\begin{equation}
J_{nuc} = \sum_{j=1,2}\left ( \Gamma_j iS_j U + U iS_j \Gamma_j \right ) \ ,
\label{nuc_cur}
\end{equation}
with $\Gamma _{j}$ denoting the $NN\eta ^{\prime }$ vertex and $S_{j}$ the nucleon 
(Feynman) propagator for nucleon $j$. The summation runs over the two interacting 
nucleons, 1 and 2. $U$ stands for the meson-exchange $NN$ potential. It is, in 
principle, identical to the potential $V$ appearing in the $NN$ scattering equation, 
except that here meson retardation effects (which are neglected in the potential 
entering in Eq.(\ref{scateqn})) are kept as given by the Feynman prescription. 

The structure of the $NN\eta ^{\prime }$ vertex, $\Gamma _{j}$, in
Eq.(\ref{nuc_cur}) is derived from the Lagrangian density
\begin{equation}
{\cal L}(x) =  - g_{NN\eta^\prime} \bar\Psi_N(x) \gamma_5 \left[
\left( i\lambda + {1-\lambda \over 2m_N}\gamma^\mu\partial_\mu \right)
\eta^\prime(x) \right] \Psi_N(x) \ ,
\label{NNps}
\end{equation}
where $g_{NN\eta ^{\prime }}$ denotes the $NN\eta ^{\prime }$ coupling constant and 
$\lambda $ is the parameter controlling the pseudoscalar(ps) - pseudovector(pv) 
admixture. $\eta ^{\prime }(x)$ and $\Psi_N(x)$ stand for the $\eta ^{\prime }$ and 
nucleon field, respectively; $m_{N} $ denotes the nucleon mass.

As mentioned in the introduction, the coupling constants $g_{NN\eta ^{\prime}}$ and
the ps-pv mixing parameter $\lambda$ are poorly known at present. The predictions 
for $g_{NN\eta ^{\prime}}$ range anywhere from 1.9 to 7.5 
\cite{Dumbrajs,Zhang,Bernard}; an estimate based on the dispersion method even gives 
$g_{NN\eta^\prime}$ consistent with zero \cite{Grein}. Zhang et al.\cite{Zhang} in 
their analysis of the photoproduction on protons of $\eta^\prime$ mesons used 
$\lambda=1$ for the ps-pv mixing parameter. Bernard et al. \cite{Bernard}, in their 
analysis of the $pp \rightarrow pp\eta^\prime$ process, extracted a value of 
$\lambda = 0.4 \pm 0.1$ in conjunction with the value of $g_{NN\eta^\prime} = 2.5 \pm  
0.7$ determined from a recent measurements of $\Delta\Sigma$ in deep inelastic 
lepton-nucleon scattering \cite{Altarelli}. The latter quantity is related to 
$g_{NN\eta ^{\prime}}$ \cite{Efremov}. SU(3) symmetry, together with the OZI rule 
\cite{OZI}, relates the $NN\eta^\prime$ coupling to the $NN\eta $ coupling: 
$g_{NN\eta ^{\prime }}=g_{NN\eta }$ tan($\alpha_{P}$), where $\alpha _{P}\equiv 
\theta _{P}-\theta _{P(ideal)}\simeq -45{{}^\circ}$ denotes the deviation from the 
pseudoscalar ideal mixing angle. With the value of $g_{NN\eta }=6.14$ used in $NN$ 
scattering analysis \cite{MHE87}, we then have $g_{NN\eta ^{\prime }}\simeq 6.1$ 
which is close to the upper end of the predicted range mentioned above. The value 
of $g_{NN\eta }=6.14$ together with the $\eta -\eta^\prime$ mixing angle of 
$\theta _{P}\simeq -9.7{{}^\circ}$, as suggested by the quadratic mass formula, and 
the $NN\pi $ coupling constant of $g_{NN\pi }=13.45$ leads to the ratio $D/F\simeq 
1.43$. This is not too far from the value of $D/F\cong 1.73$ extracted from a systematic 
analysis of semileptonic hyperon decays \cite{DFratio}. In the present work we use 
the value of $g_{NN\eta ^{\prime }}=6.1$. We shall consider both of the extreme values 
of the parameter $\lambda$ in Eq.(\ref{NNps}), i.e., $\lambda=0$ and $1$. Note that 
$\eta^\prime$ is {\em not} a Goldstone boson.  Consequently there is, {\em a priori}, 
no constraint on the ps-pv admixture.

The $NN\eta^\prime$ vertex derived from Eq.(\ref{NNps}) should be provided with an 
off-shell form factor. Following Ref.\cite{Nak2}, it is assumed to be of the form
\begin{equation}
F_N(l^2) = \frac{\Lambda_N^4} {\Lambda_N^4 + (l^2-m_N^2)^2}  \ ,
\label{formfactorN}
\end{equation}
where $l^{2}$ denotes the four-momentum squared of either the incoming or outgoing 
off-shell nucleon. We also introduce the form factor given by Eq.(\ref{formfactorN}) 
at those $NNM$ vertices appearing next to the $\eta^{\prime }$-production vertex, 
where the (intermediate) nucleon and the exchanged mesons are off their mass shell 
(see Fig.~\ref{fig1}). Therefore, the corresponding form 
factors are given by the product $F_{N}(l^{2})F_{M}(q_{M}^{2})$, where $M$ stands for 
each of the exchanged mesons between the two interacting nucleons. The form factor 
$F_{M}(q_{M}^{2})$ accounts for the off-shellness of the exchanged meson and is 
taken consistently with the $NN$ potential used for generating the T-matrix. In 
Ref.\cite{Nak2} cutoff parameters in the range of $\Lambda _{N} = 1.17-1.55 ~GeV$ 
have been used in the description of $\phi$- and $\omega$-meson productions in $pp$ 
collisions. In order to reduce the number of free parameters, here, we adopt the value 
of $\Lambda_N=1.2 ~GeV$, which is also the value adopted at the $NN^*\eta^\prime$ 
vertex in the construction of the resonance current in the next subsection.

\subsection{The Resonance Current}

Currently there are no well established nucleon resonances that decay into $N\eta^\prime$. 
In Ref.\cite{Zhang} the strong peak in 
the total cross section of the $\eta^\prime$ photoproduction off protons close to 
threshold has been attributed to the $D_{13}(2080)$ resonance. On the other hand, 
a recent multipole analysis of the $\eta^\prime$ photoproduction \cite{Plotzke} 
indicates that this reaction is dominated by an $S_{11}$ and a $P_{11}$ resonance. 
In Ref.\cite{Plotzke}, apart from the resonance dominance assumption---which probably 
leads to an overestimation of the total decay width of the resonances considered, 
it is also assumed that these resonances decay only into $p\eta^\prime$ and 
$p\pi^o$. The latter decay channel also effectively accounts for all other 
decay channels that were not considered explicitly in the analysis due to lack of 
information. Although this may not affect appreciably the description of the 
$\eta^\prime$ photoproduction reaction, the $pp \rightarrow pp\eta^\prime$ process 
can be very sensitive to such an assumption. Since the pion couples strongly to the 
nucleon, an overestimation of the $p\pi^o$ partial decay width is likely to lead to a 
considerable overestimation of the cross section due to these resonances in the 
$pp \rightarrow pp\eta^\prime$ reaction. In addition to all these uncertainties 
in the extraction of the resonance parameters, there is also the 
possibility of a threshold cusp effect, as discussed in Ref.\cite{Hoehler}, that 
might explain the observed behavior of the photoproduction cross section close to 
threshold in the absence of any resonance. All these issues require further careful 
considerations. With this background in mind we will explore the 
role of the $S_{11}(1897)$ and $P_{11}(1986)$ resonances as determined in 
Ref.\cite{Plotzke}.

The resonance current, in analogy to the nucleonic current, is written as
\begin{equation}
J_{res} = \sum_{j=1,2}\sum_{N^*}\left ( \Gamma_{jN^*\eta^\prime} iS_{N^*} U_{N^*} + 
\tilde U_{N^*} iS_{N^*} \Gamma_{jN^*\eta^\prime} \right ) \ .
\label{res_cur}
\end{equation}
Here $\Gamma _{jN^*\eta^\prime}$ stands for the $NN^*\eta ^{\prime }$ vertex 
function involving the nucleon $j$. $S_{N^*}(p) = ({\mbox{$ \not \! p$}} + m_{N^*}) / 
(p^2 - m_{N^*}^2 + im_{N^*}\Gamma_{N^*})$ is the $N^*$ resonance propagator, with 
$m_{N^*}$ and $\Gamma_{N^*}$ denoting the mass and width of the resonance, 
respectively. The summation runs over the two interacting nucleons, $j=1$ and $2$, 
and also over the resonances considered, i.e., $N^* = S_{11}(1897),~P_{11}(1986)$. 
In the above equation $U_{N^*}$ ($\tilde U_{N^*}$) stands for the 
$NN\rightarrow NN^*$ ($NN^*\rightarrow NN$) meson-exchange transition potential. 
It is given by
\begin{equation}
U_{N^*} = \Gamma_{NN^*\pi}(q_\pi) i\Delta_\pi(q_\pi^2)  \Gamma_{NN\pi}(q_\pi) \ ,
\label{transpot}
\end{equation}
where $\Delta_\pi(q_\pi^2)$ denotes the (Feynman) propagator of the exchanged pion 
with four-momentum $q_\pi$ and $\Gamma_{NN\pi}(q_\pi)$ the $NN\pi$ vertex. The 
latter is taken consistently with the $NN$ potential $V$ appearing in 
Eq.(\ref{scateqn}). We note that there is also an additional contribution to 
$U_{N^*}$ from the $\eta^\prime$ exchange in Eq.(\ref{transpot}). However this is 
negligible compared to the pion-exchange contribution.

Following Ref.\cite{Benmer}, the $NN^*\eta^{\prime }$ and $NN^*\pi$ vertices in 
Eqs.(\ref{res_cur},\ref{transpot}) are obtained from the interaction Lagrangian 
densities 
\begin{eqnarray}
{\cal L}^{(\pm)}_{NN^*\eta^\prime}(x) & = & - g_{NN^*\eta^\prime}
\bar \Psi_N(x) \left\{ \left[ i\lambda\Gamma^{(\pm)} + 
\left({1 - \lambda\over m_{N^*}\pm m_N}\right)\Gamma^{(\pm)}_\mu \partial^\mu
\right]\eta^\prime(x)\right\} \Psi_{N^*}(x) + h.c. \nonumber \\
{\cal L}^{(\pm)}_{NN^*\pi}(x) & = & - g_{NN^*\pi}
\bar \Psi_N(x) \left\{ \left[ i\lambda\Gamma^{(\pm)} + 
\left({1 - \lambda\over m_{N^*}\pm m_N}\right)\Gamma^{(\pm)}_\mu \partial^\mu
\right]\vec\tau \cdot \vec\pi(x)\right\} \Psi_{N^*}(x) + h.c. \  ,
\label{NN^*ps}
\end{eqnarray}
where $\vec\pi(x)$ and  $\Psi_{N^*}(x)$ denote the pion and nucleon resonance field, 
respectively. The upper and lower signs refer to the even($+$) and odd($-$) 
parity resonance, respectively. The operators $\Gamma^{(\pm)}$ and 
$\Gamma^{(\pm)}_\mu$ in the above equation are given by
\begin{eqnarray}
\Gamma^{(+)} = \gamma_5 & \ , & \Gamma^{(+)}_\mu = \gamma_\mu\gamma_5 \nonumber \\
\Gamma^{(-)} = 1        & \ , & \Gamma^{(-)}_\mu = \gamma_\mu \ .
\label{NRpsa}
\end{eqnarray}
The parameter $\lambda$ in Eq.(\ref{NN^*ps}) controls the admixture of the two 
types of couplings: ps ($\lambda=1$) and pv ($\lambda=0$) in the case of an even 
parity resonance and, scalar ($\lambda=1$) and vector ($\lambda=0$) in the case of 
an odd parity resonance. On-shell, both choices of the parameter $\lambda$ are 
equivalent. Hereafter the mixing parameter $\lambda$ for the $NN^*\pi$ vertex is 
fixed to be $\lambda=0$. We consider the choices $\lambda=0$ and $1$ in the 
$NN^*\eta^\prime$ vertex, however.

The coupling constants $g_{NN^*\eta^\prime}$ and $g_{NN^*\pi}$ are determined from the
extracted decay widths (and masses) of the resonances from the $\eta^\prime$ 
photoproduction on protons \cite{Plotzke},
\begin{eqnarray}
(m_{S_{11}}\ ,\ \Gamma_{S_{11}}) & = & (1.897\pm0.050^{+0.030}_{-0.002}\ ,\  
0.396\pm0.155^{+0.035}_{-0.045})\ GeV \nonumber \\
(m_{P_{11}}\ ,\ \Gamma_{P_{11}}) & = & (1.986\pm0.026^{+0.010}_{-0.030}\ ,\  
0.296\pm0.100^{+0.060}_{-0.010})\ GeV \ ,
\label{resmw}
\end{eqnarray}
with partial decay widths 
\begin{eqnarray}
\Gamma(S_{11} \rightarrow p\eta^\prime) = 0.05\Gamma_{S_{11}} & \ ,\  &  
\Gamma(S_{11} \rightarrow p\pi^o) = 0.95\Gamma_{S_{11}} \nonumber \\
\Gamma(P_{11} \rightarrow p\eta^\prime) = 0.25\Gamma_{S_{11}} & \ ,\  &  
\Gamma(P_{11} \rightarrow p\pi^o) = 0.75\Gamma_{S_{11}} \ .
\label{partdecay}
\end{eqnarray}
Using these values we obtain
\begin{eqnarray}
g_{NS_{11}\eta^\prime} = 2.9   &\ ,\ & g_{NS_{11}\pi} = 2.4 \nonumber \\
g_{NP_{11}\eta^\prime} = 11.7  &\ ,\ & g_{NP_{11}\pi} = 5.2 \ .
\label{NRpscoupl}
\end{eqnarray}
Here we assume the coupling constants to be positive. We mention in advance 
that the interference of the resonance current with the rest of the currents 
(nucleonic and mesonic currents) is practically negligible.
  
In complete analogy to the nucleonic current, we introduce the off-shell form 
factors at each vertex involved in the resonance current. We adopt the same
form factor given by Eq.(\ref{formfactorN}), with $m_N$ replaced by $m_{N^*}$
at the $NN^*\eta^\prime$ and $NN^*\pi$ vertices, in order to account for the 
off-shellness of the $N^*$ resonances. We note that Zhang et al. \cite{Zhang}
have also employed this form for the form factor at the $NN^*\eta^\prime$ 
vertex in their study of $\eta^\prime$ photoproduction. In order to account 
for the off-shellness of the exchanged pion (see Eq.(\ref{transpot})), the 
$NN^*\pi$ vertex is also multiplied by an extra form factor of $F_\pi(q_\pi^2) = 
(\Lambda_\pi^2 - m_\pi^2) / (\Lambda_\pi^2 - q_\pi^2)$, the same form used at 
the $NN\pi$ vertex in the construction of the $NN$ potential $V$ entering in 
Eq.(\ref{scateqn}), with the cutoff parameter of $\Lambda_\pi=0.8 ~GeV$. We also
use the same form factor at the $NN\pi$ vertex in Eq.(\ref{transpot}).

\subsection{The Mesonic Current}

For the meson-exchange current we consider the contribution from the 
$vv\eta^\prime$ vertex with $v$ denoting either a $\rho $ or $\omega $ meson. 
As we shall show later, this gives rise to the dominant meson-exchange current. 
The $vv\eta^\prime$ vertex required for constructing the meson-exchange current 
is derived from the Lagrangian densities
\begin{eqnarray}
{\cal L}_{\rho\rho\eta^\prime}(x) & = & - \frac{g_{\rho\rho\eta^\prime}}{2 m_\rho}
\varepsilon_{\alpha\beta\nu\mu} \partial^\alpha \vec \rho^\beta(x) \cdot
\partial^\nu \vec \rho^\mu(x) \eta^\prime(x) \nonumber \\
{\cal L}_{\omega\omega\eta^\prime}(x) & = & - \frac{g_{\omega\omega\eta^\prime}}
{2 m_\omega}\varepsilon_{\alpha\beta\nu\mu} \partial^\alpha \omega^\beta(x) 
\partial^\nu \omega^\mu(x) \eta^\prime(x)  \ ,
\label{vveta'}
\end{eqnarray}
where $\varepsilon _{\alpha \beta \nu \mu }$ is the Levi-Civita antisymmetric 
tensor with $\varepsilon _{0123}=+1$. The vector meson-exchange current is then 
given by
\begin{equation}
J_{vv\eta^\prime} = \sum_{v=\rho ,\omega}\left\{[\Gamma^\alpha_{NNv}(k_v)]_1 
iD_{\alpha\beta}(k_v)
              \Gamma^{\beta\mu}_{vv\eta^\prime}(k_v, k^\prime_v)
              iD_{\mu\nu}(k^\prime_v) [\Gamma^\nu_{NNv}(k^\prime_v)]_2 
\right\} \ ,
\label{vveta'_cur}
\end{equation}
where $D_{\alpha \beta }(k_{v })$ and $D_{\mu\nu}(k^\prime_v)$ stand for the (Feynman) 
propagators of the two exchanged vector mesons (either the $\rho $ or $\omega $ mesons as 
$v =\rho $ or $\omega $) with four-momentum $k_{v }$ and $k^\prime_v$, respectively. The 
vertices $\Gamma $ involved in the above equation are self-explanatory. The $NNv $ vertex
$\Gamma _{NNv }^{\mu}(v =\rho ,\omega )$ is taken consistently with those in the potential 
used for constructing the $NN$ T-matrix in Eq.(\ref{scateqn}).

The coupling constant $g_{vv\eta^\prime}$ is determined from a systematic analysis of the
radiative decay of pseudoscalar and vector mesons in conjunction with 
vector-meson dominance. This is done following Refs.\cite{Durso,Nak2}, with the aid of an 
effective Lagrangian with SU(3) flavor symmetry and imposition of the OZI rule. The 
parameters of this model are the angle $\alpha_V (\alpha _{P})$, which measures the 
deviation from the vector(pseudoscalar) ideal mixing angle, and the coupling constant of 
the effective SU(3) Lagrangian. They are determined from a fit to radiative decay of 
pseudoscalar and vector mesons. The parameter values determined in this way in 
Ref.\cite{Durso} (model B), however, overpredict the measured radiative decay width of 
the $\eta^\prime$ meson\cite{PDG}. Therefore, we have readjusted slightly the value of 
the coupling constant of the SU(3) Lagrangian in order to reproduce better the measured 
width. We have $\alpha _{V}\cong 3.8{{}^{\circ }}$ and $\alpha _{P}\cong -45{{}^{\circ }}$, 
as given by the quadratic mass formula, and the coupling constant of the effective SU(3) 
Lagrangian of $G=7$ in units of $1/\sqrt{m_{v }m_{v }^{\prime }}$, where $m_{v }$ and 
$m_v^\prime$ stand for the mass of the two vector-mesons involved. The sign of the 
coupling constant $G$ is consistent with the sign of the $\rho\pi\gamma$ and 
$\omega\pi\gamma$ coupling constants taken from an analysis of the pion photoproduction 
data in the $\sim 1~GeV$ energy region \cite{Garc}. With these parameter values we obtain
\begin{eqnarray}
g_{\rho\rho\eta^\prime} & = & - G \sin(\alpha_P) = 4.95 \nonumber \\
g_{\omega\omega\eta^\prime} & = & -G\left( \sqrt{2}\sin^2(\alpha_V)\cos(\alpha_P)
+ \cos^2(\alpha_V)\sin(\alpha_P) \right) = 4.90 \ .
\label{meccoupl}
\end{eqnarray}

The $vv \eta ^{\prime }$ vertex $(v =\rho ,\omega )$ in Eq.(\ref{vveta'_cur}), where the 
exchanged vector mesons are both off their mass shells, is accompanied by a form factor. 
Following Ref.\cite{Nak2}, we assume the form
\begin{equation}
F_{vv\eta^\prime}(k_v^2, {k^\prime_v}^2) = 
\left( \frac{\Lambda_v^2 - m_v^2} {\Lambda_v^2 - k_v^2} \right)
\left ( \frac{\Lambda_v^2} {\Lambda_v^2 - {k^\prime_v}^2} \right ) \ .
\label{formfactorM}
\end{equation}
It is normalized to unity at $k_{v }^{2}=m_{v }^{2}$ and $k_{v }^{\prime \text{ }2}=0$, 
consistent with the kinematics at which the value of the coupling constant 
$g_{vv\eta^\prime}$ was determined. We adopt the cutoff parameter value of $\Lambda _{v}
= 1.45 ~GeV$ as determined in Ref.\cite{Nak2} from the study of the $\omega$ and 
$\phi$ meson production in $pp$ collisions.

Another potential candidate for mesonic current is the $\sigma\eta\eta ^{\prime }$-exchange
current, whose coupling constant may be estimated from the decay width of $%
\eta ^{\prime }$ into an $\eta $ and two pions. We take the Lagrangian densities
\begin{eqnarray}
{\cal L}_{\sigma\eta\eta^\prime}(x) & = & 
{g_{\sigma\eta\eta^\prime}\over \sqrt{m_\eta m_{\eta^\prime}}}
\sigma(x)\partial_\mu\eta(x)\partial^\mu\eta^\prime(x) \nonumber \\
{\cal L}_{\sigma\pi\pi}(x) & = & 
{g_{\sigma\pi\pi}\over 2m_\pi}
\sigma(x)\partial_\mu\vec\pi(x)\cdot\partial^\mu\vec\pi(x) \ ,
\label{sigmapsps}
\end{eqnarray}
where $\sigma (x)$ and $\eta (x)$ stand for the $\sigma $ and $\eta $ meson field, 
respectively, and $m_\pi$, $m_{\eta }$ and $m_{\eta ^{\prime }}$ stand for the masses of 
the $\pi$, $\eta $ and $\eta ^{\prime }$ meson.  With the coupling constant of
$g_{\sigma \pi \pi} \sim 1.3$, extracted from Ref.\cite{Schutz} in conjunction with
the $NN\sigma$ coupling constant used in the $NN$ potential $V$, and assuming that the 
measured decay width of $\Gamma (\eta ^{\prime }\rightarrow \eta +2\pi ^{o})=42KeV$ 
\cite{PDG} is entirely due to the $\sigma $ meson intermediate state, we obtain an 
(upper) estimate of $\left| g_{\sigma \eta \eta ^{\prime }} \right| \sim 1.14$. The 
sign of this coupling constant is not fixed. We shall consider both possibilities. 

The $\sigma\eta\eta^\prime$ current reads
\begin{equation}
J_{\sigma\eta\eta^\prime} = \left\{[\Gamma_{NN\sigma}]_1
i\Delta_\sigma(q_\sigma^2)
              \Gamma_{\sigma\eta\eta^\prime}(q_\sigma , q_\eta)
              i\Delta_\eta(q_\eta^2) [\Gamma_{NN\eta}(q_\eta)]_2 
\right\} +  (1 \leftrightarrow 2) \ 
\label{sigmaetaeta'_cur}
\end{equation}
in our previously-defined notation. The vertices 
$\Gamma_{NN\sigma}$ and $\Gamma_{NN\eta}$ are taken consistently with those
in the $NN$ potential $V$ in Eq.(\ref{scateqn}). The $\sigma\eta\eta^\prime$
vertex, $\Gamma_{\sigma\eta\eta^\prime}(q_\sigma , q_\eta)$, is provided with 
the same form factor given by Eq.(\ref{formfactorM}), except for the replacement
$m_v \rightarrow m_\eta$.

The total mesonic current is then given by
\begin{equation}
J_{mec} = J_{vv\eta^\prime} + J_{\sigma\eta\eta^\prime} \ .
\label{mec_cur}
\end{equation}

There are, of course, other possible mesonic currents, such as the $\omega \phi
\eta ^{\prime }$- and $\phi \phi \eta ^{\prime }$-exchange currents, that
contribute to $\eta ^{\prime }$ meson production in $pp$ collisions.
Their contributions can be estimated in a systematic way following Ref.\cite{Nak2}
and are found to be negligible.

\section{Numerical Results and Discussion}

Once all the ingredients are specified, the total cross section for the
reaction $p+p\rightarrow p+p+\eta ^{\prime }$ can be calculated. We will
first analyze each current separately, exploring the role of individual 
contributions and the uncertainties associated with them, and then combine 
them later in the section.

The $N^*$ resonance current contribution to the total cross section as a function 
of excess energy $Q$ is shown in Fig.~\ref{fig2}. The excess energy is defined as 
$Q\equiv \sqrt{s}-\sqrt{s_{o}}$, where $\sqrt{s}$ denotes the total center-of-mass
energy of the system and $\sqrt{s_{o}} \equiv 2m_{N}+m_{\eta^{\prime }}$, its 
$\eta ^{\prime }$-production threshold energy. Here we also display the recent 
experimental data from the COSY-11 \cite{Moskal} and SPESIII \cite{Hibou} 
collaborations. The mixing parameter $\lambda$ in Eq.(\ref{NN^*ps}) for the 
$NN^*\pi$ vertices is fixed to be $\lambda=0$. Hereafter $N^*$ stands for both the 
$S_{11}(1897)$ and $P_{11}(1986)$ resonances unless otherwise indicated. The upper 
panel shows the results with $\lambda =0$ in the $NN^*\eta^\prime$ vertices. We see 
that the cross section is largely dominated by the $S_{11}(1897)$ contribution 
(dash-dotted line). The $P_{11}(1986)$ resonance leads to an energy dependence (dashed 
line) of the cross section steeper than that of the $S_{11}(1897)$ resonance, although 
in the excess energy region considered its contribution is practically negligible. The 
solid line corresponds to the total contribution, which is close to the data.
The lower panel corresponds to the choice of $\lambda=1$ at 
the $NN^*\eta^\prime$ vertices. Compared to the case of $\lambda=0$, we see that the 
mixing parameter has practically no influence on the $S_{11}(1897)$ contribution. This 
is due to the fact that this resonance is practically on its mass shell in the 
near-threshold energy region, so that both choices of the parameter $\lambda$ are 
equivalent. Although still very small compared to the $S_{11}(1897)$ contribution, the 
$P_{11}(1986)$ resonance contribution increases substantially in the low excess energy 
region compared to the $\lambda=0$ case. This is because this resonance is off-shell 
and the ps coupling mixes efficiently the positive and negative energy states. The pv 
coupling ($\lambda=0$) suppresses this mixing. The interference with the $S_{11}(1897)$ 
contribution is now destructive. In this case the total contribution (solid curve) 
is slightly reduced compared to that in the upper panel. These results are consistent 
with the recent analysis of $\eta^\prime$ photoproduction \cite{Plotzke}. We recall that 
the coupling constants at the $NN^*\eta^\prime$ and $NN^*\pi$ vertices entering in the 
resonance current have been determined from the decay widths extracted in Ref.\cite{Plotzke}. 

\begin{figure}[h]
\vspace{20cm}
\includegraphics{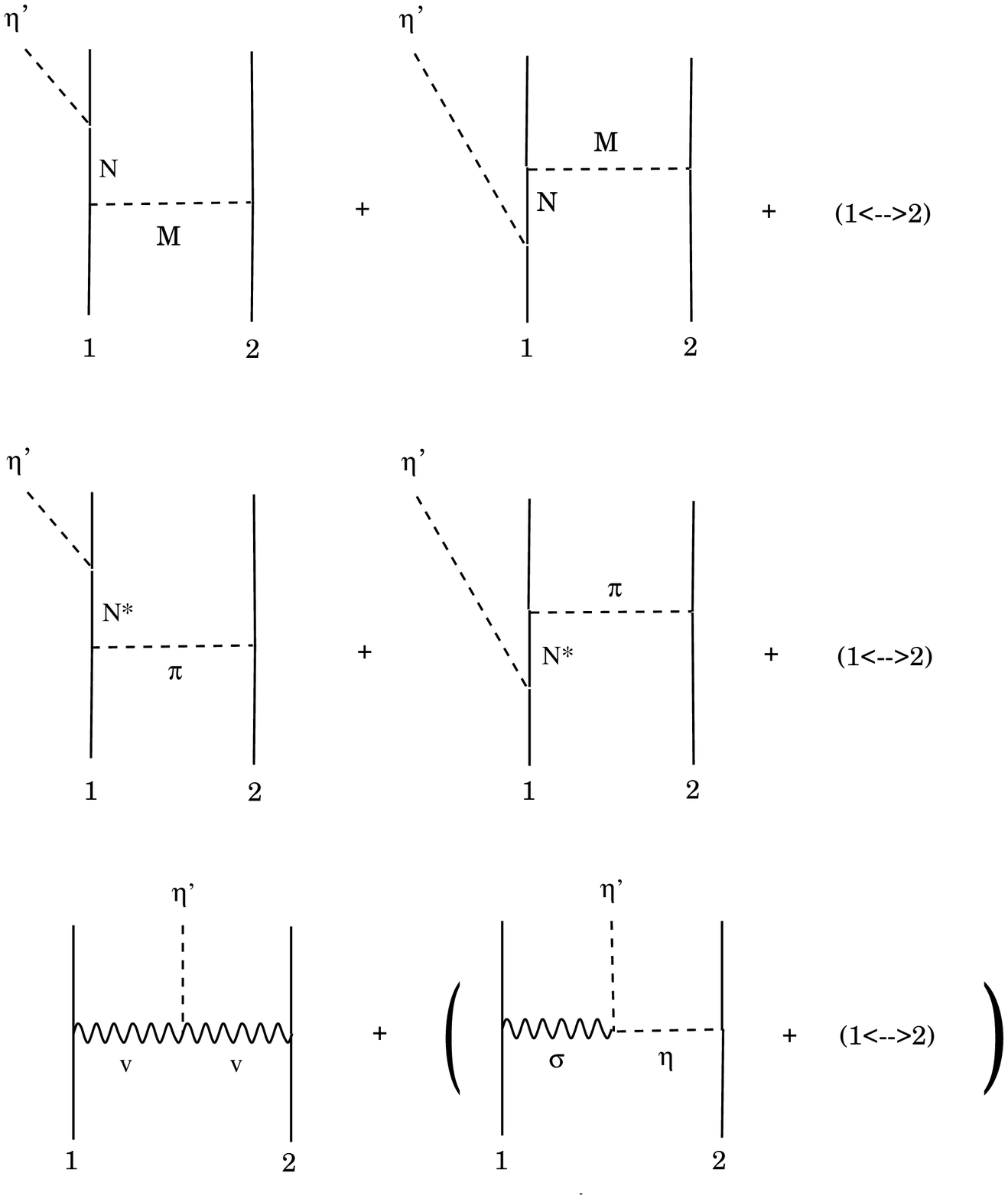}
\caption{$\eta^\prime$ meson production currents included in 
the present study. Upper row: nucleonic current $J_{nuc}$,  
$M = \pi, \eta, \rho, \omega, \sigma, a_o$.
Middle row: nucleon resonance current $J_{res}$, $N^*=S_{11}(1897),~P_{11}(1986)$.
Lower row: mesonic current $J_{mes}$, $v = \rho, \omega$.}
\label{fig1}
\end{figure}

\begin{figure}[h]
\vspace{15cm}
\includegraphics{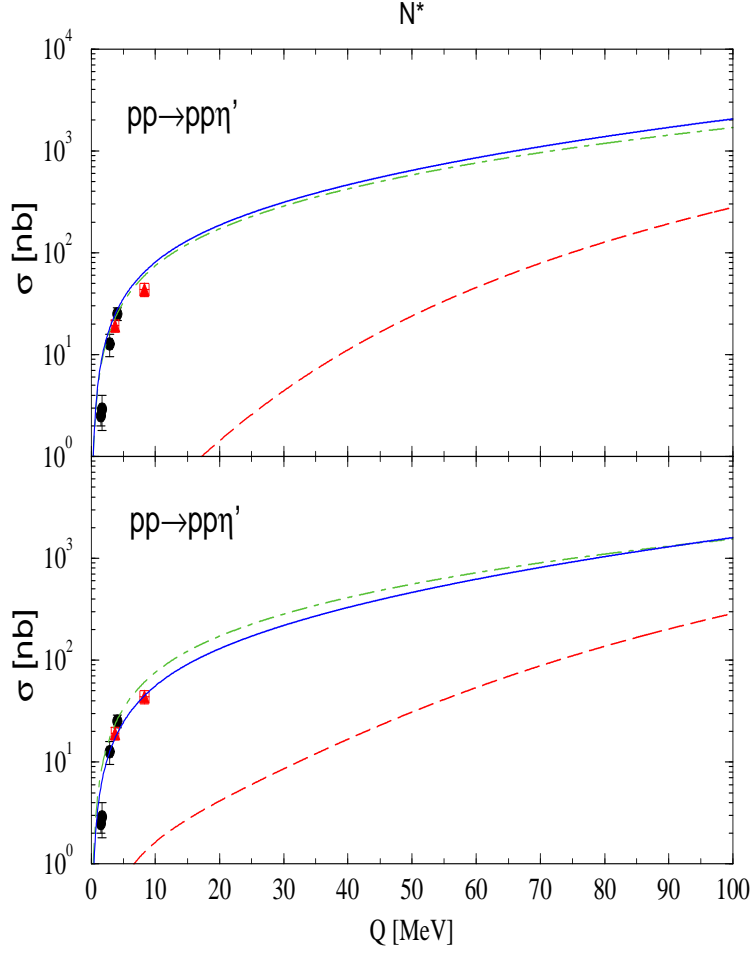}
\caption{$N^*$ resonance current contribution to the total cross section for the
reaction $pp\rightarrow pp\eta^\prime$ as a function of excess energy $Q$.
The upper panel corresponds to the results with the choice $\lambda=0$ in 
Eq.(\protect\ref{NN^*ps}) at the $NN^*\eta^\prime$ vertices, the lower 
panel to those with the choice $\lambda=1$. The dashed-dotted curves correspond 
to the $S_{11}(1897)$ resonance contribution, the dashed curves to the 
$P_{11}(1986)$ resonance contribution. The solid curves are the total contributions. 
The experimental data are from Refs.~\protect\cite{Moskal,Hibou}.} 
\label{fig2}
\end{figure}

Although we find the consistency between our result above for $pp \rightarrow 
pp\eta^\prime$ and the $\eta^\prime$ photoproduction analysis of Ref.\cite{Plotzke}, 
care must be taken in drawing any premature conclusions about the basic $\eta^\prime$ 
production mechanisms in these reactions. As mentioned in subsection III.B, given the 
assumptions made in Ref.\cite{Plotzke} in the extraction of the partial decay widths, 
the corresponding coupling constants, $g_{NN^*\eta^\prime}$ and $g_{NN^*\pi}$, are 
subject to considerable uncertainties. In particular, the resonance dominance assumption
does not allow any other possible production mechanism. As we shall show below,
the vector-meson exchange current yields also a cross section comparable to the
data. This indicates that the resonance current contribution shown in Fig.\ref{fig1} is 
probably overestimated. Another source of uncertainty in the resonance current arises 
from the off-shell form factors introduced at the $NN^*\pi$ vertex. Although the dominant 
$S_{11}(1897)$ resonance contribution is insensitive to the form factor at the 
$NN\eta^\prime$ vertex because the resonance is nearly on its mass shell near the threshold 
energy, the corresponding exchanged pion (see Fig.\ref{fig1}) is far off-shell. This 
makes the resonance contribution very sensitive to the form factor involving this pion. 
In the present calculation we have employed the same monopole form factor used at the 
$NN\pi$ vertex entering in the $NN$ potential $V$ in Eq.(\ref{scateqn}) with the cutoff 
parameter of $\Lambda_\pi=0.8 ~GeV$. One might then argue that, within the resonance 
dominance assumption, the $\gamma p \rightarrow p\eta^\prime$ and $pp \rightarrow 
pp\eta^\prime$ processes can be described consistently---provided that one uses a soft form 
factor at the $NN^*\pi$ vertex. Whatever the arguments, it is clear that further 
investigation is required in order to determine better the relevant coupling 
constants before a more definite conclusion can be drawn about the role of the nucleon 
resonance current in the $pp \rightarrow pp\eta^\prime$ reaction. Of course one should 
also keep in mind the possibility that the strong peak in the cross section observed in 
$\eta^\prime$ photo-production close to threshold may be due to a cusp effect 
\cite{Hoehler} and {\em not} due to a resonance.

\begin{figure}[h]
\vspace{10cm}
\includegraphics{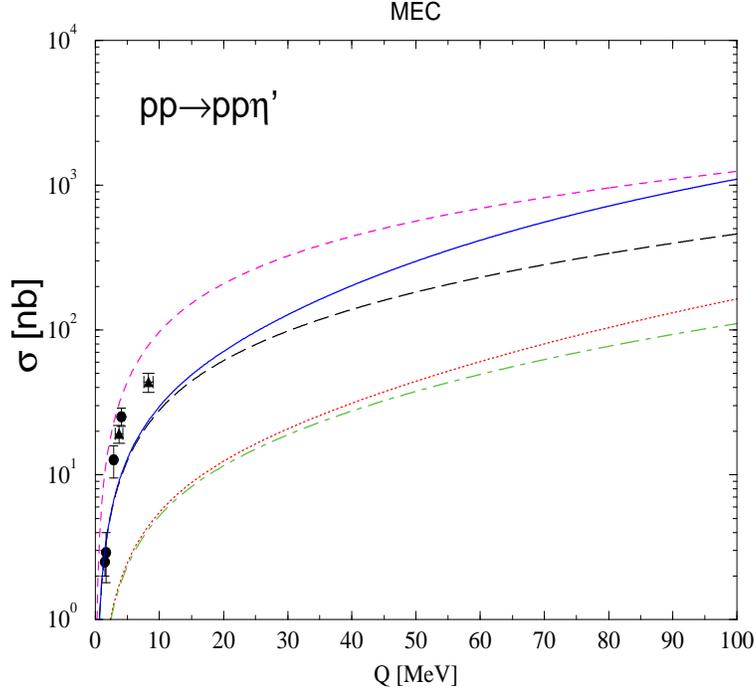}
\caption{Mesonic current contribution to the total cross section for the
reaction $pp\rightarrow pp\eta^\prime$ as a function of excess energy $Q$.
The dash-dotted curve correspond to the $\sigma\eta\eta^\prime$-exchange 
and the dotted curve to the $\omega\omega\eta^\prime$-exchange current 
contribution. The dashed curve represents the $\rho\rho\eta^\prime$-exchange
current contribution. The solid curve is the total contribution corresponding 
to choosing the negative sign for the $\sigma\eta\eta^\prime$ coupling constant,
the short-dashed curve to the positive sign. The experimental data are from 
Refs.~\protect\cite{Moskal,Hibou}.} 
\label{fig3}
\end{figure}

We turn now to investigate the mesonic current contribution, which is displayed in 
Fig.~\ref{fig3}. Among the exchange currents considered, the 
$\rho\rho\eta^\prime$ exchange gives rise to the dominant contribution (dashed line). 
The $\omega\omega\eta^\prime$- and $\sigma\eta\eta^\prime$-exchange current 
contributions (dotted and dash-dotted line, respectively) are of similar magnitude. 
However they are smaller by about a factor of five compared to the 
$\rho\rho\eta^\prime$-exchange contribution. The slightly different energy dependence 
of the cross section resulting from the $\rho\rho\eta^\prime$- and 
$\omega\omega\eta^\prime$-exchange currents is due to the tensor coupling in the 
$NN\rho$ vertex. This coupling is absent in the $NN\omega$ vertex used here. As 
mentioned in subsection III.C, the sign of the $\sigma\eta\eta^\prime$-exchange current 
relative to $vv\eta^\prime$-exchange currents ($v=\rho, \omega$) is not fixed; therefore 
we consider both possibilities. Assuming the negative sign for 
$g_{\sigma\eta\eta^\prime}$, the interference with the $vv\eta^\prime$-exchange 
current is destructive, yielding the total contribution represented by the solid 
line. If $g_{\sigma\eta\eta^\prime}$ is positive, then the interference is 
constructive and gives the total contribution shown by the dotted line. Although 
this latter choice leads to an overprediction of the data, we cannot exclude it 
at this stage because, as we shall show below, the nucleonic current can interfere 
destructively with the mesonic current so that the combined contribution may be 
smaller than that of mesonic current alone. Moreover, the off-shell form factor 
at the $\eta^\prime$-production vertices in the mesonic current is not 
well-determined at present. As mentioned in subsection III.C, we employ the same 
form factors at the $\pi\rho\omega$ and $\pi\rho\phi$ vertices as in the study of 
$\omega$- and $\phi$-meson production in $pp$ collisions \cite{Nak2}, which are 
themselves subject to considerable uncertainties. Indeed, a reduction of about 
$10\%$ in the value of the cutoff parameter can bring the short-dashed curve in 
Fig.~\ref{fig3} onto the data. Therefore neither sign of $g_{\sigma\eta\eta^\prime}$ 
can be excluded from the present analysis. In this regard, measurements of cross 
sections at higher excess energies may be useful in deciding between the two choices 
of the sign in the $\sigma\eta\eta^\prime$ coupling.

\begin{figure}[h]
\vspace{15cm}
\includegraphics{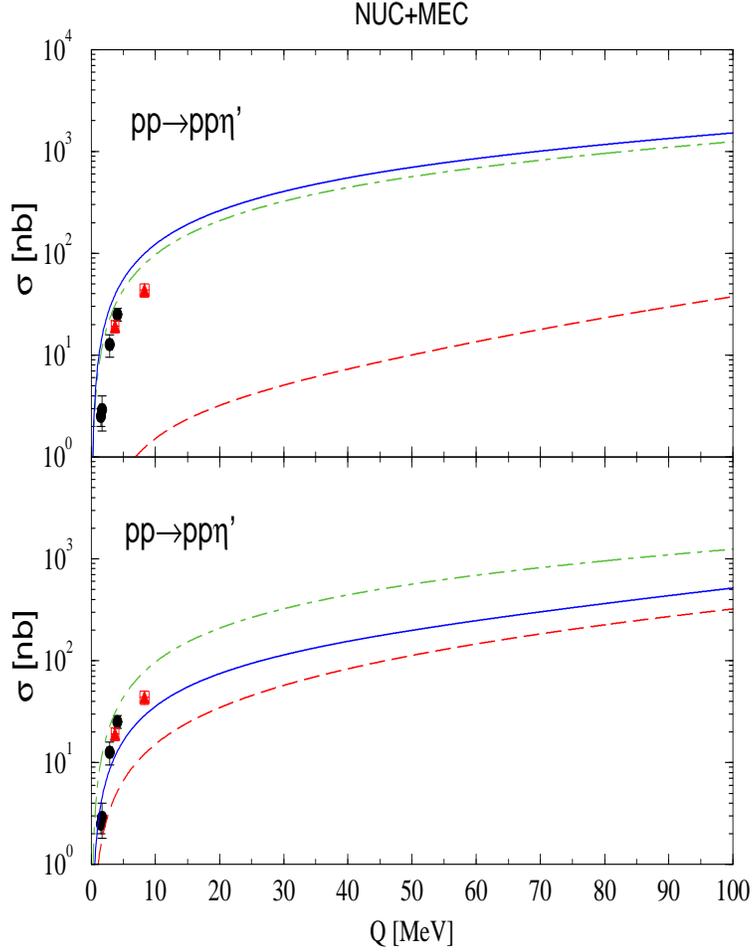}
\caption{Nucleonic-plus-mesonic current contributions to the total cross section 
for the reaction $pp\rightarrow pp\eta^\prime$ as a function of excess energy $Q$.
The upper panel corresponds to the results with the pv coupling ($\lambda=0$) at 
the $NN\eta^\prime$ vertex in the nucleonic current, the lower 
panel to those with ps coupling ($\lambda=1$). The dashed-dotted curve in both 
panels represents the total mesonic current contribution shown by the dashed curve
in Fig.~\protect\ref{fig3}. The nucleonic current contributions are represented by the 
dashed curves and the corresponding total contributions by the solid curves.
The experimental data are from Refs.~\protect\cite{Moskal,Hibou}.} 
\label{fig4}
\end{figure}

\begin{figure}[h]
\vspace{15cm}
\includegraphics{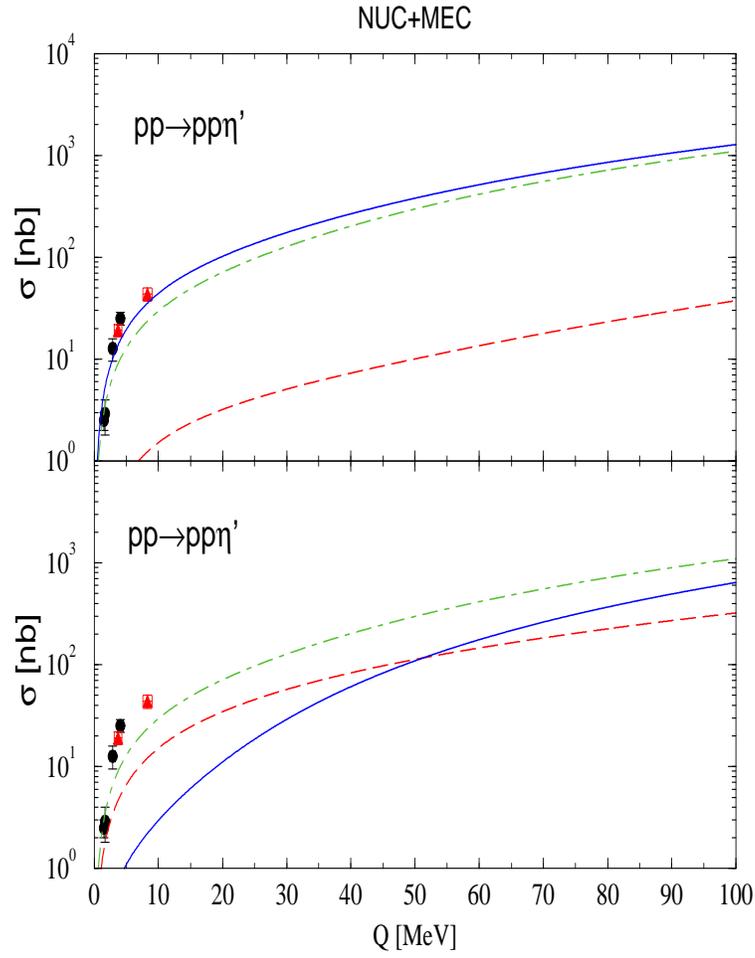}
\caption{Same as Fig.~\protect\ref{fig4}, except for the total mesonic current 
contribution, which corresponds to the result represented by the solid curve in
Fig.~\protect\ref{fig3}}. 
\label{fig4a}
\end{figure}

In the upper panel of Fig.~\ref{fig4}, the nucleonic current contribution (dashed 
curve) is shown with pv coupling at the $NN\eta^\prime$ vertex. The contribution 
from the mesonic current with the choice of positive sign for 
$g_{\sigma\eta\eta^\prime}$ (dashed curve in Fig.~\ref{fig3}) is also shown here as 
the dash-dotted curve. As we can see, the nucleonic current is about two orders of 
magnitude smaller than the mesonic current. Their interference is constructive and 
results in the total contribution shown by the solid curve, which overpredicts the data. 
The lower panel of Fig.~\ref{fig4} shows the same results, except that ps coupling is 
used at the $NN\eta^\prime$ vertex in the nucleonic current contribution. Compared to 
the case of pv coupling, we observe a considerable enhancement of the nucleonic current. 
This is due to the strong admixture of the positive and negative energy nucleon states 
provided by the ps coupling; with pv coupling this mixing is suppressed. Furthermore 
the interference with the mesonic current is now destructive, resulting in a total 
contribution (solid line) that is substantially smaller than the mesonic current 
contribution alone.  
Fig.~\ref{fig4a} shows the same results as in Fig.~\ref{fig4}, except that here the 
mesonic current contribution corresponds to the choice of negative sign for  
$g_{\sigma\eta\eta^\prime}$ (solid curve in Fig.~\ref{fig3}). We see a similar feature
to that observed in Fig.~\ref{fig4}, yielding net cross sections that underpredict 
the data. In particular, we see a strong destructive interference between the mesonic 
and nucleonic currents in the case of the ps coupling at the $NN\eta^\prime$ vertex 
(lower panel). 
We observe that the effect of the nucleonic current may be overestimated here because we 
have used a rather large $NN\eta^\prime$ coupling constant, as discussed in subsection III.A. 
In any case, given the fact that the nucleonic current is small compared to other currents, 
a quantitative determination of the $NN\eta^\prime$ coupling constant using the $pp 
\rightarrow pp\eta^\prime$ reaction is unlikely to be possible---at least not before the 
other dominant currents are better understood.

\begin{figure}[h]
\vspace{12cm}
\includegraphics{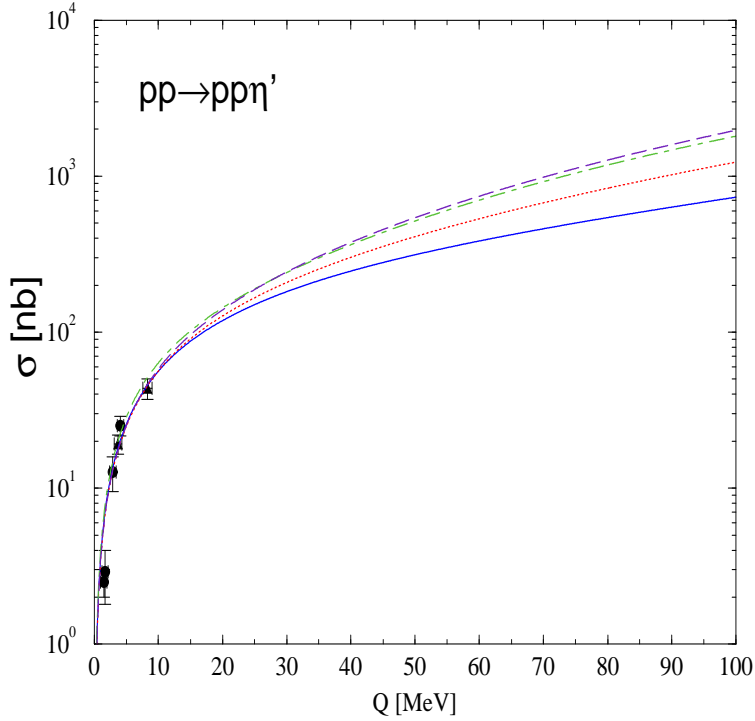}
\caption{The total cross section for the reaction $pp\rightarrow pp\eta^\prime$ 
as a function of excess energy $Q$. The solid line represents the mesonic-plus-nucleonic
current contribution corresponding to the upper panel of Fig.~\protect\ref{fig4},
except for the value of the ps-pv mixing parameter at the $NN\eta^\prime$ vertex, which 
is taken to be $\lambda=0.7$. The dotted line represents the result corresponding to the
mesonic-plus-nucleonic current shown in the lower panel of Fig.~\protect\ref{fig4}; it 
includes also the nucleon resonance current, whose product of the coupling constants 
$g_{NN^*\eta^\prime} g_{NN^*\pi}$ is multiplied by an arbitrary reduction factor of $1/2$. 
The dash-dotted and dashed curves are the same as the dotted curve, except that they 
correspond to the mesonic-plus-nucleonic current shown in the upper and lower panels of 
Fig.~\protect\ref{fig4a}, respectively. The corresponding reduction factors of the product 
$g_{NN^*\eta^\prime} g_{NN^*\pi}$ are $0.625$ and $1$. The experimental data are from 
Refs.~\protect\cite{Moskal,Hibou}.} 
\label{fig5}
\end{figure}

The above considerations expose the role of individual currents as well as the 
current uncertainties associated with them in the description of the $pp \rightarrow 
pp\eta^\prime$ reaction. We combine now all the currents. In the following we consider 
all the possibilities for the mesonic-plus-nucleonic current contribution as shown in 
Figs.~\ref{fig4} and \ref{fig4a}. For each scenario considered we adjust the coupling 
constants $g_{NN^*\eta^\prime}$ and $g_{NN^*\pi}$ in the resonance current so as to 
reproduce the data, taking $\lambda=1$ in the $NN^*\eta^\prime$ vertices. The results 
are shown in Fig.~\ref{fig5}. For the case of the mesonic-plus-nucleonic current shown 
in the upper panel of Fig.~\ref{fig4}, the corresponding resonance contribution is set 
to zero because the mesonic-plus-nucleonic current already overpredicts the data and 
the resonance current can only enhance the cross section further. We recall that the 
interference between the resonance and other currents is practically negligible. 
Therefore we have here adjusted the ps-pv mixing parameter in the $NN\eta^\prime$ vertex 
to $\lambda=0.7$ in order to reproduce the data, and the corresponding result is shown in 
the solid curve of Fig.~\ref{fig5}. For the choice of the mesonic-plus-nucleonic current 
as shown in the lower panel of Fig.~\ref{fig4}, the resonance current is required in order 
to reproduce the data. We have adjusted this current contribution by multiplying the 
product of the coupling constants $g_{NN^*\eta^\prime} g_{NN^*\pi}$ for both 
$N^*=S_{11}(1897)$ and $P_{11}(1986)$ resonances by an arbitrary reduction factor of $1/2$. 
The result is shown by the dash-dotted line. For the scenario of the mesonic-plus-nucleonic 
current shown in the upper panel of Fig.~\ref{fig4a}, the reduction factor of the product 
$g_{NN^*\eta^\prime}g_{NN^*\pi}$ required in the resonance current is $0.625$. The 
corresponding result is shown by the dotted curve. Finally, for the scenario of the lower 
panel in Fig.~\ref{fig4a}, no reduction of the resonance current is required in order to 
reproduce the data, since this corresponds to an extreme case where the destructive 
interference between the nucleonic and mesonic currents leads to a mesonic-plus-nucleonic 
current contribution that is nearly negligible. The short-dashed curve is the corresponding 
result. Fig.~\ref{fig5} reveals the different energy dependences arising from each scenario 
considered. We see that if one wants to learn about the production mechanism from the total 
cross section then one has to go to much higher energy region than that covered by the 
presently available data.

At this point we note that, in view of the latter scenario discussed in Fig.\ref{fig5} above, 
it is not surprising that the approach of Ref.\cite{Sibirtsev} describes the $pp \rightarrow 
pp\eta^\prime$ data. In Ref.\cite{Sibirtsev} the $\pi N \rightarrow \eta^\prime N$ amplitude, 
extracted from the measured cross section, has been used as the production mechanism. The 
$\pi N \rightarrow \eta^\prime N$ amplitude obtained in this way must include the nucleon 
resonance contribution that we have considered in this work, which leads to a strong $\pi N 
\rightarrow \eta^\prime N$ transtion. We emphasize, however, that one has to be careful in 
drawing any premature conclusion here. First, as discussed before, our nucleon resonance current 
is subject to considerable uncertainties. Second, although the calculation of Ref.\cite{Sibirtsev} 
is based on the measured $\pi N \rightarrow N \eta^\prime$ cross section, it requires an 
introduction of an off-shell form factor at the $NN\pi$ vertex in order to account for the 
off-shellness of the exchanged pion. As we have pointed out, the exchanged pion is far off-shell 
in the present reaction and consequently the resulting cross sections are very sensitive to the 
corresponding form factor at the $NN\pi$ vertex. We note that in Ref.\cite{Sibirtsev} a monopole 
form factor with the cutoff parameter of $\Lambda_\pi = 1.3~GeV$, as used in the Bonn $NN$ potential, 
has been employed. There are, however, a number of indications that the $NN\pi$ form factor is much 
softer ($\Lambda_\pi \sim 0.8~GeV$) than that used in the Bonn potential \cite{Lattice,Bockman}. This
would reduced the contribution of the one-pion exchange to the $pp \rightarrow pp\eta^\prime$ process 
substantially. The above consideration gives an idea of the kind of uncertanties involved in the 
theoretical predictions.

\begin{figure}[h]
\vspace{10cm}
\includegraphics{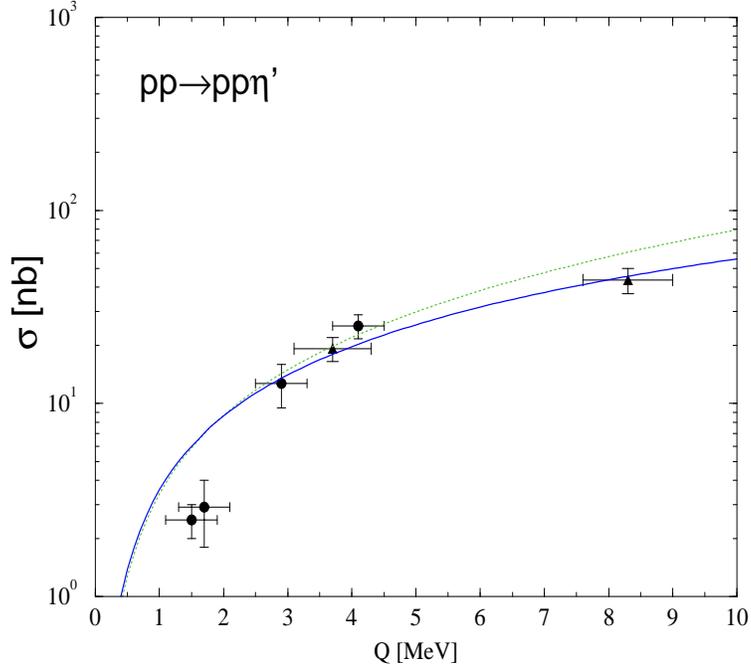}
\caption{Same as Fig.~\protect\ref{fig5} for excess energy below 10 $MeV$. Only one
total contribution (solid curve) is shown since all the curves shown in 
Fig.~\protect\ref{fig5} are almost indistinguishable at the scale used here. The 
dotted line corresponds to the result with only the phase space-plus-final state 
interaction.}
\label{fig6}
\end{figure}

We now focus our attention on the excess energy region close to threshold where
the data exist. Fig.~\ref{fig6} shows the results of Fig.~\ref{fig5} in the 
range of excess energy up to $Q=10~MeV$. Here, only one total current contribution 
(solid curve in Fig.~\ref{fig5}) is shown since all four scenarios considered 
lead to almost indistinguishable results at the scale used in the limited energy 
domain considered in the figure. We note that the energy dependence of the cross 
section is not reproduced. In particular, after adjusting the ps-pv mixing parameter 
or the reduction factor of $g_{NN^*\eta^\prime} g_{NN^*\pi}$ in order to reproduce the 
higher energy data points, as has been done in Fig.~\ref{fig5}, 
we are not able to describe the lowest two data points. We observe that the data have 
considerable uncertainties in the excess energy, presumably due to background subtraction 
that becomes more difficult for energies very close to the threshold. Nevertheless the 
results indicate the need for some mechanism(s) that introduces an additional energy 
dependence of the total cross section very close to threshold. The energy dependence of 
the total cross section near threshold is basically determined by the final state $NN$ 
interaction \cite{Watson}. This is illustrated by the dotted line in Fig.~\ref{fig6}, 
which corresponds to the result calculated taking into account only the 
three-body phase space-plus-final state interaction; it has been normalized to the solid 
curve in the low excess energy region. We see that the production currents start to modify 
the energy dependence given by the phase space-plus-final state interaction only for 
energies above $Q=4-5~MeV$. Therefore the predicted energy dependence in the near- 
threshold region is unlikely to be modified by the {\it basic} production mechanism(s). 
In fact, all the individual currents considered in this work lead practically to 
the same energy dependence in the restricted energy range considered in Fig.~\ref{fig6}.

\begin{figure}[h]
\vspace{15cm}
\includegraphics{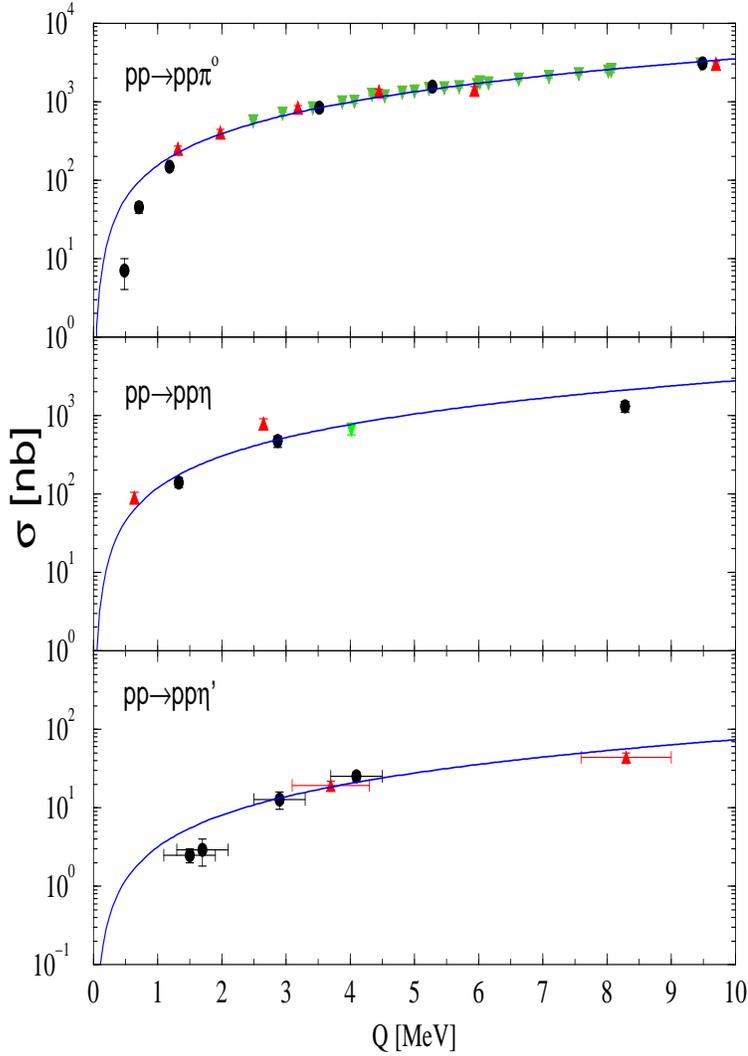}
\caption{The total cross section for $\pi^o$-, $\eta$- and $\eta^\prime$-meson 
production in $pp$ collisions as a function of excess energy $Q$. The curves represent
the corresponding energy dependence given by the phase space-plus-final state 
interaction with {\em no} Coulomb force. In the upper panel the experimental 
data are from Refs.\protect\cite{iucf90,iucf92,celsiuspi}; in the middle panel from 
Refs.\protect\cite{Hibou,spes3eta2,celsiuseta}; in the lower panel from
Refs.\protect\cite{Moskal,Hibou}.} 
\label{fig7}
\end{figure}

The fact that the energy dependence given by the three-body phase space-plus-final 
state interaction does not reproduce the observed energy dependence in $pp\rightarrow
pp\eta^\prime$ process has been a subject of attention since the publication of the 
COSY-11 data \cite{Moskal,Bernard,Baru}. In Fig.~\ref{fig7} we show the measured cross 
section data for $\pi^o$, $\eta$ and $\eta^\prime$ production in $pp$ collisions 
together with the corresponding energy dependence predicted by the phase 
space-plus-final state interaction only (solid curves). Apart from the fact that 
in $\eta^\prime$ production the deviation starts a little lower in excess energy than 
in $\pi^o$ production, we see no indication for any peculiar feature in $\eta^\prime$ 
production.

\begin{figure}[h]
\vspace{10cm}
\includegraphics{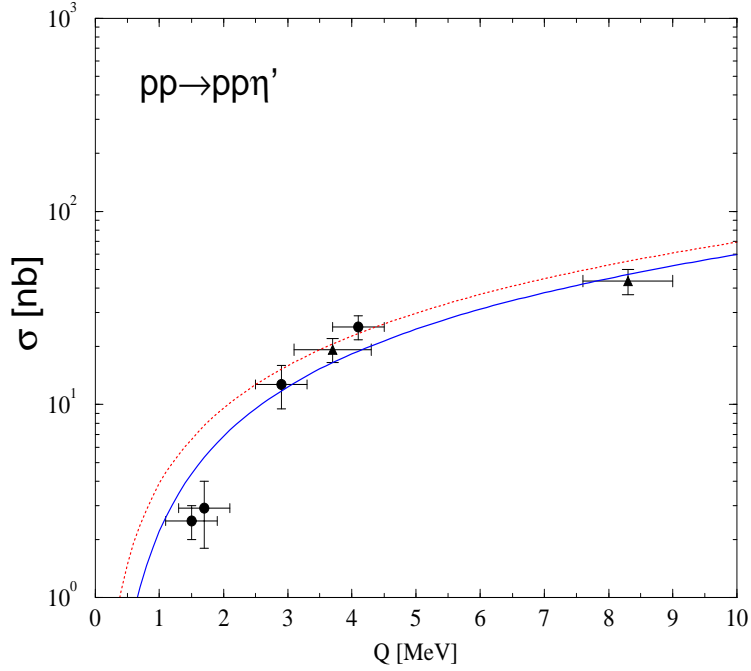}
\caption{The total cross section for the reaction $pp\rightarrow pp\eta^\prime$ 
as a function of excess energy $Q$. The solid curve corresponds to the total current
contribution including the Coulomb force in the final state interaction.
The dotted line corresponds to that when the Coulomb force is switched off. 
Here the Paris T-matrix has been used as the final state interaction.
The experimental data are from Refs.~\protect\cite{Moskal} (circle) and 
\protect\cite{Hibou} (triangle).} 
\label{fig8}
\end{figure}

The most trivial source of influence on  the energy dependence of the cross section 
close to threshold is the Coulomb force \cite{Miller}, which has not been considered so 
far in our calculations. We follow the Gell-Mann and Goldberger two-potential formalism 
\cite{GellMann} to include the Coulomb force in the final state interaction. The 
perturbed wave function is calculated exactly by solving the Schr\"odinger equation in 
coordinate space with the nuclear-plus-Coulomb potential. For the nuclear potential 
we use the Paris potential \cite{Lacomb}. The half-off-shell (pure) Coulomb T-matrix 
is obtained in analytic form \cite{Dilin}.
We note that care must be taken in order to use consistently a $NN$ T-matrix 
constructed within a non-relativistic formulation---such as the Paris T-matrix---in a 
relativistic approach. We have followed Ref.\cite{Herrmann} in order to use the Paris 
T-matrix in the present relativistic approach. Also, in the following, the production 
current has been kept the same as that used in conjunction with the Bonn B potential, 
except for a readjustment of the reduction factor of the product $g_{NN^*\eta^\prime} 
g_{NN^*\pi}$ in the nucleon resonance current that was needed to reproduce the data. 
We are aware that by using the Paris potential instead of the Bonn B potential we may 
loose the consistency between the production current and the final state interaction, 
however the 
energy dependence of the cross section is practically the same---when the Coulomb force 
is switched off---as that given by the Bonn B potential over a wide range of excess energy. 
The result of including the Coulomb effect as described above is shown in Fig.~\ref{fig8} 
by the solid line. The dotted line corresponds to the result when the Coulomb force is 
switched off. As can be seen, the Coulomb effect is considerable; in particular, it reduces 
the cross section by almost $40\%$ at $Q=1.5~MeV$.

The picture that emerges from the above considerations is that there is no 
obvious indication of a need for other mechanisms than those already 
considered here in order adequately to describe the existing data. 
A close comparison of the COSY-11 \cite{Moskal} and SPESIII \cite{Hibou} data, however, 
seems to indicate different trends in the energy dependence. In this connection, new 
measurements at higher energies, and even remeasurement of some of the data  
near $Q\sim 4~MeV$, are extremely important for resolution of this issue. The new data at 
higher energies will certainly impose more stringent constraints on the parameters 
of our model. They would also tell us whether a proper description of the $pp \rightarrow 
pp\eta^\prime$ requires the inclusion of the $p\eta^\prime$ final state interaction, as has 
been argued recently~\cite{Baru}. The $p\eta^\prime$ final state interaction as discussed 
in Ref.\cite{Baru} can be viewed as the higher order terms in the production current 
treated in the present work. In fact, the $\eta ^{\prime }$-emission current defined in 
the previous section (cf. Eqs.(\ref{nuc_cur},\ref{res_cur},\ref{mec_cur})) is, in part, 
just the Born term of a more general current that can be obtained by using the T-matrix 
amplitude for the $M+p\rightarrow \eta^\prime +p$ transition, where $M$ denotes any meson 
of interest. This can easily be seen if we disregard the nucleon labeled 2 in Fig.~\ref{fig1}; 
the current then becomes nothing other than the Born term of the $M+p\rightarrow \eta^\prime+p$ 
transition amplitude. The T-matrix amplitude for the $M+p\rightarrow \eta ^{\prime }+p$ 
process may be separated into the so-called pole and non-pole terms according to Pearce 
and Afnan \cite{Pearce}. By using the physical baryon mass and physical $NB\eta^\prime$ 
($B=N,N^*$) and $NNM$ vertices in the baryonic current, the pole term of the $M+p
\rightarrow \eta^\prime+p$ T-matrix amplitude is fully accounted for in the present work. 
What is taken in the Born approximation is, therefore, the non-pole part of the T-matrix 
only.

\section{Summary}

We have investigated the reaction $pp \rightarrow pp\eta^\prime$ within a 
relativistic meson-exchange model of hadronic interactions. We find that the 
$S_{11}(1897)$ resonance as determined in a recent multipole analysis of
$\eta^\prime$ photoproduction \cite{Plotzke} gives rise to a contribution that
reproduces the measured cross section and allows for no other production 
mechanisms. This is consistent with the analysis of Ref.\cite{Plotzke}, where 
the nucleon resonance parameters have been extracted under the resonance dominance 
assumption. Whether this is already the entire physics of the production mechanism is an
open question that requires further investigation. Indeed, in Ref.\cite{Plotzke}, in 
addition to the assumption of resonance dominance, only the $p\eta^\prime$ and $p\pi^o$ decay 
channels have been explicitly considered; the $p\pi^o$ decay channel accounts for 
{\em all} possible decay channels 
other than the $p\eta^\prime$. The coupling constants $NN^*\eta^\prime$ and 
$NN^*\pi$ obtained from the decay widths extracted under these assumptions are, 
therefore, subject to considerable uncertainties which, in turn, lead to 
corresponding uncertainties in the resonance current contribution to the cross 
section. A more accurate determination of the relevant resonance parameters is 
called for before a more definite conclusion can be drawn about the role of the 
$N^*$ resonances in the $pp \rightarrow pp\eta^\prime$ reaction. We should also 
keep in mind that the strong peak in the cross section close to threshold in 
$\eta^\prime$ photoproduction may be due to a cusp effect \cite{Hoehler}, and not
to the resonance assumed in Ref.\cite{Plotzke}. All these 
issues associated with the resonance require further careful investigation.   

The mesonic current gives rise to cross sections which are comparable to the 
measured values. It is dominated by the $\rho\rho\eta^\prime$-exchange current. 
Depending on the choice of the sign of the $\sigma\eta\eta^\prime$ coupling constant, 
the calculated cross sections overpredict or underpredict the data. However neither 
choice can be excluded at present---mainly due to uncertainties associated with the 
baryonic current. Data in higher energy regions than currently available may provide 
the necessary constraint to fix the sign of this coupling, as the different signs lead 
to different energy dependences.   

Since the nucleonic current is found to be small, it is unlikely that the $pp 
\rightarrow pp\eta^\prime$ reaction can be used quantitatively to determine the 
$NN\eta^\prime$ coupling constant---at least not before the other dominant currents 
are under better control.

As expected from earlier calculations \cite{Miller}, the Coulomb force in the final state 
interaction is found to play a crucial role in explaining the observed energy dependence 
of the cross section close to threshold. The basic production mechanisms start to 
influence the energy dependence of the cross section only for excess energies above 
$\sim 5~MeV$. For lower energies the energy dependence is determined by the three-body 
phase space-plus-final state interaction \cite{Watson}. New data at higher energies will 
certainly impose constraints on the production mechanisms. In addition 
to more exclusive observables than the total cross section, a combined analysis of
production processes using both electromagnetic and hadronic probes should 
help to disentangle the different production mechanisms.

{\bf Acknowledgment}
We are grateful to J. Haidenbauer and S. Krewald for useful discussions. We also thank 
Ulf Mei\ss ner and J. Haidenbauer for critical readings of the manuscript.
H.F.A. acknowledges the partial support from FONDECYT, grant No. 1970508.

\vfill \eject

\vglue 0.5cm

\newpage
\vglue 0.5cm

\newpage
\vglue 2cm

\newpage
\vglue 1cm

\newpage
\vglue 1cm

\newpage
\vglue 1cm

\newpage
\vglue 2cm

\newpage
\vglue 1cm

\newpage
\vglue 2cm

\end{document}